\documentclass[sigplan,screen,10pt]{acmart}

\setcopyright{none}
\acmPrice{}
\acmDOI{}
\acmYear{2018}
\copyrightyear{2018}
\acmISBN{}
\acmConference[]{Technical Report}{2018}{NUS}

\renewcommand{\ttdefault}{lmtt}

\usepackage{amsmath}
\usepackage{amsthm}
\usepackage[caption=false]{subfig}
\usepackage{pifont}
\usepackage{pgfplots}
\usepackage{tikz}
\usetikzlibrary{calc}
\usetikzlibrary{patterns}
\usepackage{listings}
\usepackage{lstlinebgrd}
\usepackage{etoolbox}
\usepackage{fancyvrb}
\usepackage{multirow}
\usepackage{enumitem}

\theoremstyle{definition}

\newtheorem{example}{Example}

\patchcmd{\quote}{\rightmargin}{\leftmargin 2ex \rightmargin}{}{}

\newcommand{\ignore}[1]{}

\newcommand{\cmark}[0]{\ding{51}}
\newcommand{\xmark}[0]{\ding{55}}

\newcommand{\textttz}[1]{\texttt{#1}}

\newcommand{\lowfat}[0]{LowFat~}

\newcommand{\base}[1]{\mathit{base}(#1)}

\definecolor{mygreen}{rgb}{0,0.5,0}

\newcommand{\authnote}[2]{{\bf \textcolor{blue}{#1}: \em \textcolor{red}{#2}}}
\renewcommand{\authnote}[2]{}

\setcitestyle{sort&compress}

\begin{document}

\title[An Extended Low Fat Allocator API]{An Extended Low Fat Allocator API and Applications}
\subtitle{}

\author{Gregory J. Duck}
\affiliation{%
    \department{Department of Computer Science}
        \institution{National University of Singapore}
}
\email{gregory@comp.nus.edu.sg}
\author{Roland H. C. Yap}
\affiliation{%
    \department{Department of Computer Science}
        \institution{National University of Singapore}
}
\email{ryap@comp.nus.edu.sg}

\begin{abstract}
The primary function of memory allocators is to allocate and deallocate
chunks of memory primarily through the \texttt{malloc} API. 
Many memory allocators also implement other API extensions, such as
deriving the size of an allocated object from the object's pointer,
or calculating the base address of an allocation from an
interior pointer.
In this paper, we propose a general purpose extended allocator API built
around these common extensions.
We argue that such extended APIs have many applications and demonstrate
several use cases,
such as (manual) memory error detection,
meta data storage,
typed pointers
and compact data-structures.
Because most existing allocators were not designed for the extended API,
traditional implementations are expensive or not possible.

Recently, the \lowfat allocator for heap and stack objects has been
developed.
The \lowfat allocator is an implementation of the idea of low-fat pointers,
where object bounds information (size and base) are encoded into the native
machine pointer representation itself.
The ``killer app'' for low-fat pointers is automated bounds
check instrumentation for program hardening and bug detection.
However, the \lowfat allocator can also be used to implement highly optimized
version of the extended allocator API, 
which makes the new applications (listed above) possible.
In this paper, we implement and evaluate several applications based
efficient memory allocator API extensions using low-fat 
pointers.
We also extend the \lowfat allocator to cover
global objects for the first time.
\end{abstract}

\maketitle

\section{Introduction}\label{sec:intro}

Memory allocators are used heavily in languages without garbage
collection, for example, in \verb_C_/\verb_C++_.
Memory allocation (and deallocation), canonically this is through
\verb|malloc|/\verb|free| (or \verb_C++_'s \verb_new_ operators),
is well understood and studied \cite{wilson-survey}.
There are many widely used memory allocators, to name a few,
the Lea \cite{lea-malloc}, jemalloc \cite{je-malloc}, and
TCMalloc \cite{tcmalloc}.
Most allocators provide APIs for allocating (\verb|malloc| and friends)
and deallocation (\verb|free| and friends). For brevity,
we will simply call this the {\em malloc API}.

The nub of the malloc API has remained fairly static for a long time, focusing
on the core functionality of allocation and deallocation of memory.
However, there is other functionality which can be offered, 
separate from the main allocation and deallocation tasks.
Indeed, some allocators provide some extended non-core functionality, and
we argue that extensions,
such as returning 
information about allocated objects, is both useful and can support 
a variety of applications. 
While some allocators have some non-core malloc API 
extensions, we propose a unifying set of malloc API extensions.

Our extensions leverage the \lowfat allocator which has been
recently developed for efficient bounds checking
\cite{duck16heap,duck17stack}.
The \lowfat allocator allows for certain operations, such as calculating
the allocation size/base/offset of pointers very efficiently, which
forms the foundation of our API extensions.
This is important for applications where the extended API is heavily used,
e.g., in bounds
checking potentially every read/write can make use of \lowfat
operations.
Our API extension also allows for uniform treatment of all objects
(globals, stack and heap), in contrast, traditional memory allocators
only provide APIs for heap objects.
Although some similar APIs already exist --- e.g.,
the Boehm conservative garbage collector \cite{boehm88garbage}
also provides some similar functionality since the garbage collector
also needs some of the operations we propose ---
by exploiting the properties of \lowfat pointers, our implementation
is very efficient, with many operations implementable in a few inlined
low-latency instructions.
While low-fat pointers have been implemented for heap~\cite{duck16heap}
and stack~\cite{duck17stack} objects,
in this paper we also extend low-fat pointers to also cover global
objects, thereby covering all three main object kinds.

We show how to apply the extended malloc API to several applications,
including:
(manual) memory error checking, 
efficient and general meta data storage and retrieval,
typed pointers, and 
compact data-structures.
For each application we provide some (mini)benchmarks to support our claims.
Berger et al. \cite{Berger2001} propose the need for composable
memory allocators, here, we argue the case for applications which
leverage new functionality beyond memory allocation/deallocation.

In summary, the main contributions of this paper are the following:
\begin{itemize}
\item \underline{\emph{Low-fat Globals}}:
    In addition to heap and stack objects,
    we extend low-fat pointers to also cover global objects
    for the first time.
    This means that low-fat pointers are now applicable to all
    three main object kinds: heap, stack and globals.
\item \underline{\emph{An Extended \lowfat Allocator API}}:
    We present an extended version of the \verb+malloc+ API which gives
    additional operations outside the core allocation functionality.
    The extended API leverages low-fat pointers which allows for
	very efficient implementation of key operations.
\item \underline{\emph{Applications}}:
    We present several
    novel applications, made possible by the extended malloc API, for
	non-traditional use cases, including: manual memory error checking;
	hidden meta-data; typed/tagged pointers; and compact vectors.
	We also evaluate the applications to show that they are 
	efficient either from a time or space perspective.
\end{itemize}

The paper is organized as follows:
Section~\ref{sec:alloc} summarizes the existing \lowfat allocator
for heap and stack objects, and then we present a novel extension for
low-fat global objects.
We also evaluate the performance of the \lowfat allocator against some
more established competitors.
Section~\ref{sec:api} presents the \lowfat allocator extended API, as
well as details the efficient implementation of each operation.
Finally, in Section~\ref{sec:applications}, we present and evaluate
several applications of the extended \lowfat allocator API.

\section{LowFat Allocation Design and Implementation}\label{sec:alloc}

This section describes the \lowfat allocator's design and implementation.
In a memory allocator, the precise system details can be important.
Throughout this paper, we will tailor the implementation details
for the \verb|x86_64| architecture and Linux operating system.

\subsection{Background: Low-fat Pointers}

Low-fat pointers~\cite{duck16heap, duck17stack, kwon13lowfat} are a method
for encoding object \emph{bounds information} (object's size and base)
into the native machine pointer representation itself.
For example, a highly simplified low-fat pointer encoding may be implemented
as follows:
\begin{Verbatim}[samepage=true]
union { void *ptr;
        struct {uintptr_t size:10;  // MSB
        uintptr_t unused:54; } meta;} p;
\end{Verbatim}
Here the object size is represented explicitly as a field \verb+size+, and
the base address can be encoded implicitly by ensuring object's are
aligned to an address that is a multiple of \verb+size+, thus
$\mathit{base}(p) = p - (p \bmod p\texttt{.size})$.
Crucially we see that a low-fat pointer is the same size as
a machine pointer, i.e. (\verb+sizeof(p) == sizeof(void *)+).
Low-fat pointers generally require a machine architecture with sufficient
bit-width, i.e., 48 or 64bit pointers, such as the \verb+x86_64+.

\begin{figure}
\begin{center}
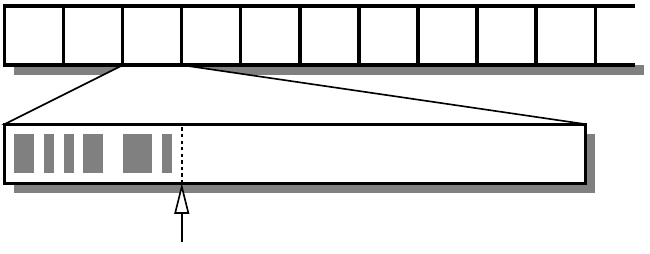
\vspace{-1em}
\end{center}
\caption{\lowfat memory layout.\label{fig:pool}}
\end{figure}

This simplified low-fat pointer encoding is difficult to implement
in practice as it imposes strong constraints
on the program's virtual address space layout.
Instead we focus on the \emph{flexible} low-fat pointer encoding
of~\cite{duck16heap, duck17stack}, which we shall refer to as \emph{\lowfat}.
The general idea of \lowfat is to partition the program's virtual
address space into several large equally-sized \emph{regions}.
There are two main types of regions:
\emph{low-fat regions} which contain objects managed by the \lowfat allocator,
and \emph{non-fat regions} that contain everything else.
In \cite{duck16heap}, region \#0 is non-fat, and we will also follow
that approach.
The basic idea is that each low-fat region will
service allocations of a given size range, as illustrated in
Figure~\ref{fig:pool}.
For example, region \#1 handle allocations of sizes 1-16 bytes,
region \#2 handles sizes 17-32 bytes, region \#3 33-48 bytes, etc.
The mapping between sizes and low-fat regions is called the \emph{size
configuration}~\cite{duck16heap}, represented by a sequence $\mathit{Sizes}$.
For example, the $\mathit{Sizes}$ for~\cite{duck17stack} is as follows:
\begin{align*}
    \mathit{Sizes} = \langle 16, 32, 48, 64, 80, 96, 112, 128, 144, .. \rangle
\end{align*}
Generally, the size configuration should have the following
properties:
\begin{enumerate}
\item \label{prop:16} All sizes must be a multiple of 16bytes;
\item \label{prop:pow2} $\mathit{Sizes}$ must include a power-of-two
    sub-sequence, i.e.: \\
    $\mathit{Sizes} \cup \langle 16, 32, 64, 128, 256, .. \rangle = 
     \mathit{Sizes}$; and
\item \label{prop:large} Large multi-page sizes should be powers-of-two,
    i.e.: \\
    $\mathit{Sizes} \cup \langle 16\mathit{KB}, 32\mathit{KB}, 64\mathit{KB},
        .. \rangle = \mathit{Sizes}$; and
\end{enumerate}
Property~\ref{prop:16} ensures the allocator obeys the default alignment
of standard \verb+malloc+ for 64-bit systems.
Property~\ref{prop:pow2} is needed to support both the stack and
global low-fat pointers (discussed below) as well as support for the
\texttt{memalign} API.
Property~\ref{prop:large} keeps $|\mathit{Sizes}|$ compact, since large
multi-page objects can be ``rounded-up'' to the nearest power-of-two multiple
without wasting memory (the ``padding'' will remain virtual).
Note that properties~\ref{prop:16}, \ref{prop:pow2} and \ref{prop:large}
are consistent with each other.
The full low-fat allocator parameters used in this paper are listed
in Appendix~\ref{app:params}.

During allocation, an object of $\mathit{size}$ is rounded-up to the next
allocation size
($\mathit{allocSize} \geq \mathit{size}$) that fits, which some
caveats discussed below.
For the object $O$ to qualify as \emph{low-fat}, two main properties must be
satisfied:
\begin{itemize}
\item \textbf{Region}: The object $O$ is allocated from the sub-heap in
    region \#I, where $\mathit{Sizes}[I] = \mathit{allocSz}$; and
\item \textbf{Alignment}: The object $O$ is $\mathit{allocSz}$-aligned.
\end{itemize}
These two properties ensure that the object's size and base address
can be quickly calculated from a (possibly interior) pointer to the
object $O$.
This will be elaborated on in Section~\ref{sec:api}.

Memory for the low-fat regions is created during program initialization,
e.g., as a \texttt{preinit\_array} callback.
Regions do not grow or shrink during
program execution, rather, the initial size is assumed to be large enough to
accommodate all ``reasonable'' future memory requirements of the program.
For example, the implementation of~\cite{duck17stack} assumes a region size
of 32$\mathit{GB}$.
The low-fat regions are initially \emph{virtual memory} reserved using \texttt{mmap} using
the \texttt{NORESERVE} flag, and thus does not initially consume any
RAM/swap resources.
Memory resources are only consumed for the parts of each region that are actually
allocated and used by the program.
Finally, each region is further partitioned into three heap/stack/global
\emph{sub-regions} to handle allocations of the corresponding memory type
(see Figure~\ref{fig:pool}).
This will be discussed further below.

\subsection{\lowfat Heap Allocation}

The exact memory allocation algorithm used for heap objects within each region
is left open.
The \cite{duck16heap, duck17stack} implementation uses a simple free-list
allocator design that partitions the heap sub-region into
\emph{used} and \emph{unused} space.
Objects in the \emph{used} space are either allocated and in use by the
program, or have been freed and placed on a ``free list'' awaiting
reallocation.
When a call to \verb+lowfat_malloc(s)+ occurs, the \lowfat allocator:
\begin{enumerate}
\item Determines which region \#$i$ corresponding to size \verb+s+ should the
      allocation be serviced from; and
\item Pops an entry from the free-list for region \#$i$ if non-empty; 
      else
\item Allocate a new object from the \emph{unused} space otherwise.
\end{enumerate}
Calls to \verb+free(p)+ are handled by pushing the allocated space pointed to
by \verb+p+ onto the corresponding free-list.
For large objects, it is sometimes necessary to return free'ed memory back to
the operating system, which is done using the 
\verb+madvise+ system call with the \verb+DONTNEED+ flag.

Since all allocations of a particular size class are serviced from a single
region,
this has the side-effect of simplifying the overall allocator design.
For example, merging of adjacent free objects is disallowed, thus the
corresponding logic to do so is not needed by the allocator.
The trade-off is that this may lead to more fragmented memory since free'ed
objects can only be reallocated as objects within the same size class.
On the other hand, since the allocation size can be determined from the
pointer (i.e., which region does the pointer point to?),
and since there is no need to implement adjacent free object merging, the
\lowfat allocator also eliminates the need to store an explicit
``malloc header'', meaning that objects are tightly packed.
In contrast, the standard \verb+stdlib+ \verb+malloc+ implementation
for Linux appends a 16byte header to every object.
That said, we highlight that in this paper, the main aim of the \lowfat 
allocator is to support an enriched \lowfat allocator API 
presented in Section~\ref{sec:api}, rather than to design an allocator that directly
competes with the current state-of-the-art on performance.

\subsubsection*{Benchmarking the \lowfat Heap Allocator}

\begin{figure*}
\pgfplotsset{
    axisA/.style={
        ybar=0pt,
        ymin=0,
        ymax=500,
        xticklabels={
           \texttt{\scriptsize perlbench},
           \texttt{\scriptsize bzip2},
           \texttt{\scriptsize gcc},
           \texttt{\scriptsize mcf},
           \texttt{\scriptsize gobmk},
           \texttt{\scriptsize hmmer},
           \texttt{\scriptsize sjeng},
           \texttt{\scriptsize libquantum},
           \texttt{\scriptsize h264ref},
           \texttt{\scriptsize omnetpp},
           \texttt{\scriptsize astar},
           \texttt{\scriptsize xalancbmk},
           \texttt{\scriptsize milc},
           \texttt{\scriptsize namd},
           \texttt{\scriptsize dealII},
           \texttt{\scriptsize soplex},
           \texttt{\scriptsize povray},
           \texttt{\scriptsize lbm},
           \texttt{\scriptsize sphinx3}
        },
        ytick={100,200,300,400,500},
        yticklabels={
            {\scriptsize 100s},
            {\scriptsize 200s},
            {\scriptsize 300s},
            {\scriptsize 400s},
            {\scriptsize 500s},
        },
        bar width=3.5pt,
        x tick label style={rotate=45,anchor=east,yshift=-2.5pt,xshift=3pt},
        width=0.88\textwidth,
        height=5cm,
        xtick=data,
        xtick pos=left,
        ytick pos=left,
        legend cell align={left},
        legend pos=north east,
        major tick length=0.08cm,
        enlarge x limits={true, abs value=0.8},
        grid style={gray!20},
        grid=both,
        legend image code/.code={%
            \draw[#1] (0cm,-0.1cm) rectangle (3pt,3pt);
        },
        legend pos=outer north east,
        title style={yshift=-6pt}
        }
}
\pgfplotstableread{
0   262  270  256  271     
1   433  430  422  439
2   244  398  278  263
3   202  205  205  208
4   399  402  401  410
5   326  335  279  335
6   411  409  408  404
7   280  283  285  282
8   478  486  479  477
9   224  186  176  227
10  328  317  312  356
11  174  138  136  178
12  344  355  344  352
13  304  304  304  310
14  240  273  234  244
15  172  171  166  177
16  126  124  123  127
17  230  220  215  232
18  421  415  407  413
}\dataset
\begin{tikzpicture}
    \begin{axis}[
        axisA
      ]
      \addplot[fill=red!20] table[x index=0,y index=1] \dataset;
      \addplot[fill=red]  table[x index=0,y index=2] \dataset;
      \addplot[pattern color=red, pattern=north east lines]  table[x index=0,y index=3] \dataset;
      \addplot[pattern color=red, pattern=north west lines]  table[x index=0,y index=4] \dataset;
      \legend{{\scriptsize {\tt stdlib}},{\scriptsize \lowfat},
        {\scriptsize {\tt jemalloc}}, {\scriptsize Boehm}};
    \end{axis}
\end{tikzpicture}
\caption{Evaluation of the \lowfat and other heap allocators against the
    SPEC2006 benchmark suite.\label{fig:mallocs}}
\end{figure*}
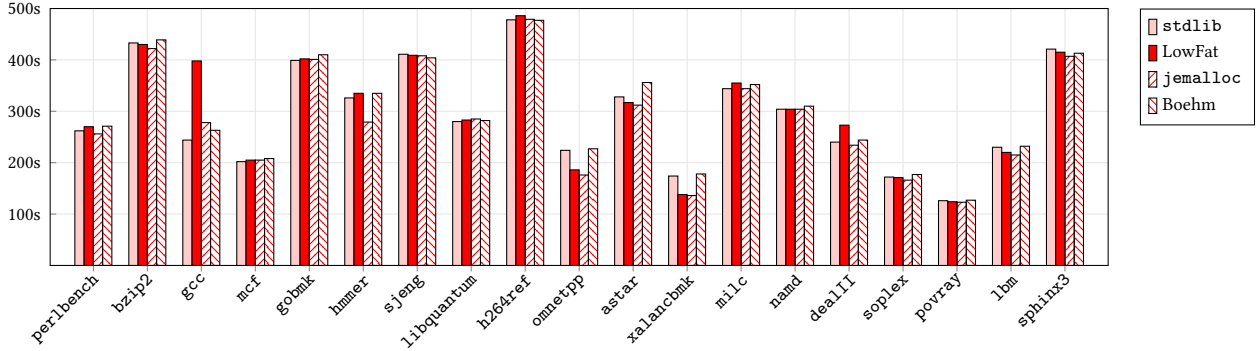

We present some benchmarks to evaluate the performance of the \lowfat
allocator against some more established alternatives.
All experiments (including in later sections) are run on a Xeon E5-2630v4
processor (clocked at 2.20GHz) with 32GB of RAM on Linux.
The compiler used is LLVM 4.0.0 at {\tt -O2}, and we
evaluate against the SPEC2006 benchmark suite.
We compare the \lowfat implementation of~\cite{lowfat_github} against
\verb+stdlib+ \verb+malloc+,
\verb+jemalloc+~\cite{je-malloc}, and
the Boehm \verb+malloc+ (in manual memory management
mode)~\cite{boehm88garbage}.
The results on the SPEC2006 benchmark suite are shown in 
Figure~\ref{fig:mallocs}.
The geometric mean for \verb+stdlib+ \verb+malloc+ is 277.8 (100\%),
\lowfat is 280.9 (101.1\%), \verb+jemalloc+ is 266.8 (96.0\%), and
Boehm is 283.6 (102.1\%).

Overall we see that the \lowfat allocator is competitive against the
alternatives.
The \lowfat allocator described in this paper is intended to be a basic
prototype without the many optimizations used in mature memory allocators,
so we expect higher overheads compared to
more optimized memory allocators such as \verb+jemalloc+.
Furthermore, the \lowfat allocator is a relatively young system, meaning that
further optimizations may be implemented in the future.
We also highlight that only the \lowfat allocator supports the optimized
\lowfat API, which is the main focus of this paper.
The memory overhead for the \lowfat allocator is $\sim$3\% compared to
\verb_stdlib_ \verb_malloc_~\cite{duck16heap, duck17stack}.

\subsection{\lowfat Stack Allocation}

A \lowfat allocator for stack memory is presented in~\cite{duck17stack},
which we briefly summarize here.
The low-fat stack allocator works by maintaining a linear mapping
between the stack sub-regions (Figure~\ref{fig:pool}) and the main program
stack.
When the program requests a stack allocation of $\mathit{size}$, the
\lowfat stack allocator performs the following steps:
\begin{enumerate}
\item Round-up $\mathit{size}$ to the nearest power-of-two allocation
    size ($\mathit{allocSize}$) that fits;
\item Mask the stack pointer \verb+%rsp+ with $\mathit{allocSize}-1$.
    This $\mathit{allocSize}$-aligns \verb+%rsp+;
\item Decrement \verb+%rsp+ by $\mathit{allocSize}$, allocating space;
\item Map \verb+%rsp+ to a pointer $\mathit{ptr}$ to the stack sub-region
    corresponding to $\mathit{allocSize}$ using the linear mapping.
    The $\mathit{ptr}$ now points to the newly allocated low-fat stack
    object.
\end{enumerate}
This mapping is implemented as a compiler transformation~\cite{lowfat_github}.
Power-of-two sizes are used since this simplifies object alignment at
the cost of increased space overheads.
Stack deallocation is handled the same as before, i.e., by restoring
\verb+%rsp+ to some previous value.
The \lowfat stack allocation method is similar to the notion of
\emph{parallel shadow stacks}~\cite{dang15shadow},
but with multiple shadow stacks (one for each sub-region) and some additional
steps $\mathit{allocSize}$-aligning objects.
Having multiple shadow stacks may waste memory, however, this can be
mitigated by mapping each shadow stack to the same physical memory.
See the \emph{memory aliasing} optimization from ~\cite{duck17stack}.

\subsection{\lowfat Global Allocation}\label{sec:globals}

Previous work on low-fat pointers are restricted to
heap~\cite{duck16heap} and stack~\cite{duck17stack} objects only.
In this paper, we present an extension of \lowfat to also cover
global objects.
The basic idea is to statically allocate global objects from the global
sub-region for the corresponding allocation size.
To achieve this, we use a program transformation which
annotates global objects using a \verb+section+ attribute and then uses
a special linker script to control the location of objects.
Namely, objects of given $\mathit{size}$ are annotated with a
\begin{Verbatim}[commandchars=\\\{\},codes={\catcode`$=3\catcode`^=7}]
    attribute(section("lowfat_region_$\mathit{idx}$"))
\end{Verbatim}
\verb+section+ attribute, where $\mathit{idx}$ corresponds to the
region index for the global object's $\mathit{size}$.
The static location of the objects can then be controled via an
appropriate \emph{linker script} (\verb|ld|), e.g.:
\begin{Verbatim}[samepage,commandchars=\\\{\},codes={\catcode`$=3\catcode`^=7}]
    . = \textrm{(\emph{global sub-region \#1 address})}
    lowfat_region_1 :
    {
        KEEP(*(lowfat_region_1))
    }
    ...
\end{Verbatim}
In addition to location, alignment of global objects is controlled using the
\verb+aligned+ attribute.
Due to the power-of-two limitation of the  \verb+aligned+ attribute,
global objects are placed into the
nearest power-of-two sized region that fits
(as is the case with stack objects).

There are some (compiler tool-chain) caveats for generating global low-fat
pointers.
Firstly, the \emph{dynamic linker} does not support the \verb+section+
directive meaning that dynamically linked globals (e.g., from shared objects)
will not be low-fat pointers.
This does not affect program behavior but limits the applicability of the
\lowfat API for such objects.
The second caveat is that the compiler may assume all global objects occupy the
first 4$\mathit{GB}$ of the virtual address space.
This allows the compiler to generate slightly faster code for the \verb+x86_64+
architecture.
This assumption is violated by global low-fat pointers, meaning that
the program must be compiled using the (\verb+-mcmodel=large+)
option which disables the assumption.
The final caveat this that, like the \lowfat stack allocator, global
objects are not low-fat by default unless 
the compiler transformations described in this section are employed.

\section{\lowfat Allocator API}\label{sec:api}

The core motivation for the allocator design is to support the \lowfat
memory API, as summarized in Figure~\ref{fig:api}.
It is divided into three classes. 
Class I refers to the (traditional) \verb+malloc+ API.
The focus of this paper will be on classes II and III detailed below.

\begin{figure}
{\small
\begin{center}
\begin{tabular}{|c||l||c|c|}
\cline{2-4}
\multicolumn{1}{c||}{} & \emph{Operation} & \emph{Inlined?} & \emph{Related?} \\
\hline
\hline
\multirow{4}{*}{I} & \texttt{lowfat\_malloc} & \xmark & \cmark \\
 & \texttt{lowfat\_realloc} & \xmark & \cmark \\
 & \texttt{lowfat\_free} & \xmark & \cmark \\
 & $\cdots$ & $\cdots$ & $\cdots$ \\
\hline
\hline
\multirow{4}{*}{II} & \texttt{lowfat\_is\_ptr} & \cmark & N.A. \\
& \texttt{lowfat\_is\_heap\_ptr} & \cmark  & \xmark \\
& \texttt{lowfat\_is\_stack\_ptr} & \cmark  & \xmark \\
& \texttt{lowfat\_is\_global\_ptr} & \cmark  & \xmark \\
\hline
\multirow{5}{*}{III} & \texttt{lowfat\_index} & \cmark  & N.A. \\
& \texttt{lowfat\_size} & \cmark   & \cmark$^*$\\
& \texttt{lowfat\_base} & \cmark   & Boehm \\
& \texttt{lowfat\_offset} & \cmark & Boehm$^\dagger$ \\
& \texttt{lowfat\_usable\_size} & \cmark & Boehm$^\dagger$ \\
\hline
\end{tabular}
\end{center}
}
\caption{Summary of the \lowfat API.
Here (\emph{Inlined?}) indicates whether the operation can be inlined, and
(\emph{Related?}) indicates whether a operation is implemented by some other
related malloc API.
The caveat ($*$) means implemented with the limitation that the pointer must
be a base pointer, and ($\dagger$) means operation is not implemented directly,
but can be implemented using the API with minimal effort.\label{fig:api}}
\end{figure}

\subsection{Standard allocator functionality}
Our \lowfat allocator supports standard replacements for
\texttt{libc}'s memory allocation functions (Figure \ref{fig:api} class I), 
such as,
\verb+malloc+, \verb+free+, \verb+realloc+, \verb+memalign+, etc.
The \lowfat replacements are also aliased to versions prefixed by
``\verb+lowfat_+'', e.g.  \verb+lowfat_malloc+, etc.

Stack and global objects can be transformed automatically as a compiler
pass (e.g., as used by~\cite{duck17stack}).
As such, stack and global support is optional, and programmers may
opt not to use it.

\subsection{Core \lowfat functionality}

The motivation behind \lowfat allocation is that allows for some key
pointer operations to be implemented efficiently, namely, calculating
the size, base, offset, etc., of a pointer \verb+p+ with respect to
the original allocation.
We highlight that the operations take only a few machine instructions
making them suitable for inlining which helps efficiency and
compiler optimizations.
Since these operations are not traditionally supported by the malloc API, we refer to
these operations as the extended memory allocation API.

By design, unlike the malloc API,
these operations work uniformly,
regardless of whether the pointer is for a heap,
stack, global, interior or exterior,
just as long as the pointer is lowfat as per Section~\ref{sec:alloc}.
In Section~\ref{sec:applications}, we will describe some applications of 
the API.

Given the memory layout of Figure~\ref{fig:pool}, we can define a
fundamental operation, \verb+lowfat_index+, that maps a pointer
\verb+ptr+ to the \emph{region index} to which \verb+ptr+ belongs, as
follows:
\begin{Verbatim}[fontsize=\small,commandchars=\\\{\},codes={\catcode`$=3\catcode`^=7}]
     lowfat_index($\mathit{ptr}$) = $\mathit{ptr}$ / LOWFAT_REGION_SIZE
\end{Verbatim}
Here \verb+LOWFAT_REGION_SIZE+ is the region size and is assumed to be
a power-of-two.
For example, our reference implementation assumes \verb+LOWFAT_REGION_SIZE+
is 32$\mathit{GB}$.
Crucially, the \verb+lowfat_index+ is fast, compiling down into a
single \verb+x86_64+ shift instruction with this default:
\begin{verbatim}
    shrq $35,%rax       /* 2^35 = 32GB */
\end{verbatim}

\subsubsection*{Size (\texttt{lowfat\_size})}

One common memory allocator API operation is to determine the size of the
allocation based on a pointer to an object.
This exists in the form of \verb+malloc_usable_size+ for \verb+stdlib+'s
\verb+malloc+, \verb+HeapSize+ for the Window's \verb+HeapAlloc+,
and \verb+GC_size+ for the Boehm collector, amongst others.
Note that all of these functions assume a pointer to the base of the
allocated object.
Furthermore, such extensions typically differ on whether the size returned
accounts for any additional bytes of padding that may have been added by
the allocator.
For example, \verb+malloc_usable_size+ returns the size including the
padding, whereas \verb+HeapSize+ returns the original requested allocation
size, depending on the version of Windows.

We define \verb+lowfat_size+ to return the allocation size of
a pointer including any padding, similar to \texttt{malloc\_\-usable\_\-size}:
\begin{Verbatim}[fontsize=\small,commandchars=\\\{\},codes={\catcode`$=3\catcode`^=7}]
lowfat_size($\mathit{ptr}$) = LOWFAT_SIZES[lowfat_index($\mathit{ptr}$)]
\end{Verbatim}
Here, \verb+LOWFAT_SIZES+ is a constant lookup table mapping
region indices to the allocation sizes according to the
\emph{size configuration} defined in Section~\ref{sec:alloc}.
For region indices $i$ that are \emph{not} associated with \lowfat allocation,
we define:
\begin{Verbatim}[commandchars=\\\{\},codes={\catcode`$=3\catcode`^=7}]
    LOWFAT_SIZES[i] = SIZE_MAX
\end{Verbatim}
This definition simplifies some applications relating to bounds checking.

Note that, unlike related allocators, the \verb+lowfat_size+ works for
any interior pointer and does not assume the base address.
The other advantage is that the \verb+lowfat_size+ compiles down into
two \verb+x86_64+ instructions, one shift for \verb+lowfat_index+
followed by a memory read:
\begin{verbatim}
    movq LOWFAT_SIZES(,%rax,8),%rbx
\end{verbatim}

\newbox\memcpybox
\begin{lrbox}{\memcpybox}
\newcommand{\highlightex}[1]{%
    \ifboolexpr{
        test {\ifnumcomp{#1}{=}{3}} or
        test {\ifnumcomp{#1}{=}{4}} or
        test {\ifnumcomp{#1}{=}{5}} or
        test {\ifnumcomp{#1}{=}{6}} or
        test {\ifnumcomp{#1}{=}{10}} or
        test {\ifnumcomp{#1}{=}{11}} or
        test {\ifnumcomp{#1}{=}{12}} or
        test {\ifnumcomp{#1}{=}{13}}
    }{\color{black!20}}{}
}
\begin{lstlisting}[
    language=C,
    numbers=left,
    numberstyle=\tiny,
    basicstyle=\scriptsize\ttfamily\fontseries{m}\selectfont,
    keywordstyle=\ttfamily\fontseries{b}\selectfont,
    mathescape,
    linebackgroundcolor=\highlightex{\value{lstnumber}},
    linewidth=0.39\textwidth]
void memcpy(void *dst, void *src, int n)
{
  void *dst_base = lowfat_base(dst);
  size_t dst_size = lowfat_size(dst);
  void *src_base = lowfat_base(src);
  size_t src_size = lowfat_size(src);
  for (int i = 0; i < n; i++) {
    void *dst_tmp = dst + i;
    void *src_tmp = src + i;
    if (isOOB(dst_tmp, dst_base, dst_size))
        error();
    if (isOOB(src_tmp, src_base, src_size))
        error();
    *dst_tmp = *src_tmp;
  }
}
\end{lstlisting}
\end{lrbox}

\newbox\memcpymbox
\begin{lrbox}{\memcpymbox}
\newcommand{\highlightex}[1]{%
    \ifboolexpr{
        test {\ifnumcomp{#1}{=}{3}} or
        test {\ifnumcomp{#1}{=}{4}} or
        test {\ifnumcomp{#1}{=}{5}} or
        test {\ifnumcomp{#1}{=}{6}} or
        test {\ifnumcomp{#1}{=}{7}} or
        test {\ifnumcomp{#1}{=}{8}}
    }{\color{black!20}}{}
}
\begin{lstlisting}[
    language=C,
    numbers=left,
    numberstyle=\tiny,
    basicstyle=\scriptsize\ttfamily\fontseries{m}\selectfont,
    keywordstyle=\ttfamily\fontseries{b}\selectfont,
    mathescape,
    linebackgroundcolor=\highlightex{\value{lstnumber}},
    linewidth=0.39\textwidth]
void memcpy(void *dst, void *src, int n)
{
  size_t dst_size = lowfat_usable_size(dst);
  if (n > dst_size)
      error();
  size_t src_size = lowfat_usable_size(src);
  if (n > src_size)
      error();
  for (int i = 0; i < n; i++)
    dst[i] = src[i];
}
\end{lstlisting}
\end{lrbox}

\begin{figure*}[t]
\subfloat[Automatically instrumented (see~\cite{duck17stack}).\label{fig:memcpy}]{
\begin{minipage}{0.42\textwidth}
\usebox\memcpybox
\end{minipage}
}
~~~~ ~~~~ ~~~~ ~~~~ 
\subfloat[Optimally instrumented.\label{fig:memcpy2}]{
\begin{minipage}{0.42\textwidth}
\usebox\memcpymbox
\end{minipage}
}
\caption{Two bounds-check instrumented variants of (simple) \texttt{memcpy}.
         The instrumentation is highlighted.}
\end{figure*}

\subsubsection*{Base (\texttt{lowfat\_base}) and
    offset (\texttt{lowfat\_offset})}

Given a pointer $\mathit{ptr}$ to an allocated object $O$ of $\mathit{size}$,
then
\[
    \{\mathit{ptr}+1, .., \mathit{ptr}+\mathit{size}\}
\]
are the \emph{interior pointers} of $O$, and
$\mathit{ptr}$ is the \emph{base pointer} (a.k.a. exterior pointer) of $O$.
We can map any (possibly interior) pointer $\mathit{ptr}' \in I$ to object
$O$ to the base pointer $\mathit{ptr}$ using the following operation:
\begin{Verbatim}[commandchars=\\\{\},codes={\catcode`$=3\catcode`^=7}]
lowfat_base($\mathit{ptr}$) =
    ($\mathit{ptr}$ / lowfat_size($\mathit{ptr}$)) * lowfat_size($\mathit{ptr}$)
\end{Verbatim}
This assumes 64bit integer arithmetic, and is also equivalent to
$\mathit{ptr} - \mathit{ptr}~\%~\textttz{lowfat\_size}(\mathit{ptr})$.
This also relies on the \lowfat allocator ensuring that all allocated objects
are \emph{size}-aligned.
Assuming the pointer is stored in register \verb+%rax+ (and is an implicit
argument), and the allocation size in \verb+%rbx+, then
the \verb+lowfat_base+ operation reduces to two instructions:
\begin{verbatim}
    divq %rbx
    imulq %rbx
\end{verbatim}
    
As noted in~\cite{duck16heap}, the 64bit \verb+divq+ operation is relatively
slow (high throughput and latency~\cite{intel}), 
which may not be desirable.
There are two main approaches to optimizing \verb+lowfat_base+, namely:
\begin{enumerate}
\item Use a power-of-two-only size configuration; or
\item Use fixed-point or floating-point arithmetic.
\end{enumerate}
The first allows for the slow division to be replaced by a fast bitmask
operation, for example:
\begin{Verbatim}[commandchars=\\\{\},codes={\catcode`$=3\catcode`^=7}]
lowfat_base($\mathit{ptr}$) =
    $\mathit{ptr}$ & LOWFAT_MASKS[lowfat_index($\mathit{ptr}$)]
\end{Verbatim}
where $\textttz{LOWFAT\_MASKS}[i]$ is defined to be
$(\textttz{LOWFAT\_SIZES}[i]-1)$ for
low-fat region \#i, or $0$ otherwise.
The main disadvantage with this approach is that object sizes are rounded to
the nearest power-of-two, which leads to increased space overheads.
An alternative approach is to use \emph{fixed-point arithmetic} by defining: 
\begin{Verbatim}[fontsize=\small,commandchars=\\\{\},codes={\catcode`$=3\catcode`^=7}]
LOWFAT_MAGICS[i] = ((1 << $R$) / LOWFAT_SIZES[i]) + 1
\end{Verbatim}
for low-fat region \#i, or $0$ otherwise.
The (\verb_+1_) term is for error control,
see~\cite{duck16heap} Section 5.1.1.
Here $R$ defines the position of the radix point.
This approach and allows for a more efficient implementation
of the base operation that effectively turns a slow division operation
into a fast(er) multiplication:
\begin{Verbatim}[fontsize=\small,commandchars=\\\{\},codes={\catcode`$=3\catcode`^=7}]
lowfat_base($\mathit{ptr}$) =
  ((($\mathit{ptr}$ * LOWFAT\_MAGICS[lowfat_index($\mathit{ptr}$)])) >> $R$)
      * lowfat_size($\mathit{ptr}$)
\end{Verbatim}
A good value for $R$ is $64$, as this takes advantage of the
\verb+x86_64+'s 128bit integer multiplier, meaning the
$R$-right shift operation ``compiles away'' into a mere register renaming.
It is also possible to use floating-point arithmetic, which is more
intuitive, by defining: 
\begin{Verbatim}[commandchars=\\\{\},codes={\catcode`$=3\catcode`^=7}]
  LOWFAT_MAGICS[i] = (1.0 / LOWFAT_SIZES[i])
\end{Verbatim}
However, fixed-point avoid conversions to-and-from floating point numbers
so is generally more efficient.
The main disadvantage of fixed/floating point arithmetic is that calculations
may be affected by precision errors, which mainly affect large allocations.
Using the (\verb_+1_) term for error control, precision errors will only affect
pointers to ``near the end'' of these large allocations.
This problem is mitigated by modifying the allocator to take precision errors
into account~\cite{duck16heap}.
Namely, if an object of a given $\mathit{size}$ is potentially affected by a
precision errors in region~\#$i$, then the allocator will instead service
the allocation from the next (larger) region~\#$(i+1)$.
The maximum possible precision error for each region is calculated in
advance~\cite{duck16heap}.

The Boehm conservative garbage collector \cite{boehm88garbage} also supports \verb+GC_base+ (equivalent to our \verb+lowfat_base+)
as an $O(1)$ operation. 
However, the Boehm implementation is slower and larger (in terms of code size)
to that of low-fat pointers.
Due to the larger code size, the Boehm \verb+GC_base+ operation generally
cannot be inlined.

Finally, we define a \verb+lowfat_offset+ operation that returns
the difference from the current pointer and the base:
\begin{Verbatim}[commandchars=\\\{\},codes={\catcode`$=3\catcode`^=7}]
lowfat_offset($\mathit{ptr}$) = $\mathit{ptr}$ - lowfat_base($\mathit{ptr}$)
\end{Verbatim}
It is also possible to implement the \verb+lowfat_offset+ directly with
fixed-point arithmetic, i.e., by multiplying the fixed-point mantissa by
the allocation size.
However, since the mantissa represents the least significant bits,
a fixed-point implementation of \verb+lowfat_offset+ is impractical
due to precision errors.

\subsubsection*{Usable size (\texttt{lowfat\_usable\_size})}

Recall that the \verb+lowfat_size+ returns the allocation size for
the base or any interior pointer to the object, and this
size is the same regardless of the pointer's offset.
For many applications, we wish to know how many bytes are left
until we reach the end of the allocated space.
For this we define:
\begin{Verbatim}[commandchars=\\\{\},codes={\catcode`$=3\catcode`^=7}]
lowfat_usable_size($\mathit{ptr}$) =
    lowfat_size($\mathit{ptr}$) - lowfat_offset($\mathit{ptr}$)
\end{Verbatim}
For example, given a pointer $p$ into a buffer \verb+buf+, then the
\verb+lowfat_usable_size+ operation can determine how many bytes
are left inside \verb+buf+ from $p$ until a buffer overflow occurs.

\subsubsection*{Tests (\texttt{lowfat\_is\_ptr}, $\cdots$, \texttt{lowfat\_is\_global\_ptr})}

It is sometimes useful to test whether a pointer is low-fat or not.
The motivation is to allow inter-operation with non low-fat pointers, possibly,
from other memory allocators.
It can also be useful to test whether or not the pointer is
a low-fat heap/stack/global pointer.
These operations reduce to simple range tests, e.g.:
\begin{Verbatim}[commandchars=\\\{\},codes={\catcode`$=3\catcode`^=7}]
lowfat_is_ptr($\mathit{ptr}$) =
    ($\mathit{ptr}$ >= &$(\mathit{region} \#1)$) && ($\mathit{ptr}$ < &$(\mathit{region} \#M{+}1))$
\end{Verbatim}
Here $1..M$ ($M$ is the last region) are the indices of the low-fat regions.
The test starts from region \#1 as region \#0 is non-fat as per \cite{duck16heap}.
The narrower
\texttt{heap}/\texttt{stack}/\texttt{global} variants additionally test
which sub-region (see Figure~\ref{fig:pool}) the pointer points to.

\section{Applications}\label{sec:applications}

The \lowfat allocator implementation supports efficient implementations of
some operations.
This enables some applications that would
otherwise be too slow for other memory management systems.
In this section we explore examples of such applications, including:
manual bounds checking,
hidden meta-data,
typed pointers
and compact data-structure representations.

\subsection{Detecting Memory Errors}

Automated bounds check instrumentation is  the ``killer app''
for low-fat pointers, and this idea has been explored by previous
literature~\cite{duck16heap,duck17stack}.
The basic idea is to instrument all pointer arithmetic and memory access
with an explicit bounds check 
(\verb+isOOB+) defined as follows:
\begin{align}
        (p < \mathit{base})~\texttt{||}~
        (p > \mathit{base}{+}\mathit{size}{-}\textttz{sizeof}(\texttt{*}p))
        \tag{\texttt{isOOB}}
\end{align}
Automatic bounds instrumentation follows the schema introduced
in~\cite{duck16heap}.
The basic idea is as follows:
for all \emph{input} pointers $q$ (function arguments, return values,
or pointer values read from memory), we calculate the bounds
\emph{meta information} by calling the \verb+lowfat_size+/\verb+lowfat_base+
operations.
For example:
\begin{Verbatim}[commandchars=\\\{\},codes={\catcode`$=3\catcode`^=7}]
void f(int *$q$) \{
    void  *q_base = lowfat_base($q$);
    size_t q_size = lowfat_size($q$); ...
\end{Verbatim}
Next, for all pointers $p$ \emph{derived} from an input pointer $q$ through
pointer arithmetic ($p = q \texttt{+} k$) or
field access ($p = \texttt{\&}q\texttt{->}\mathtt{field}$),
we instrument any access to $p$ with an (\verb+isOOB+) check.
For example:
\begin{Verbatim}[commandchars=\\\{\},codes={\catcode`$=3\catcode`^=7}]
    int *p = q + k;
    if (isOOB(p, q_base, q_size)) error();
    x = *p;     {\rm\em or}     *p = x;
\end{Verbatim}
Such bounds-check instrumentation is implemented as a LLVM~\cite{llvm}
compiler pass, see~\cite{lowfat_github}.

An automatically instrumented version of a (simple) implementation of
\verb+memcpy+ is shown in Figure~\ref{fig:memcpy} (based
off~\cite{duck17stack} Figure 2).
Here the instrumented lines are highlighted, including the bounds meta
data calculation using \verb+lowfat_size+/\verb+low+\-\verb+fat_base+ shown in
lines 3--6.
Automated bounds checking has an overhead of 64\% for heap/stack/global
objects~\cite{lowfat_github}, although lower overheads are possible
depending on what optimizations are enabled
(generally trading error coverage for speed).

\subsubsection*{Manual Bounds Checking}

Automatic bounds instrumentation has the advantage in that
it requires minimal intervention on behalf of the programmer
(e.g., changing the compiler's flags).
However, the automatically generated instrumentation is generally sub-optimal.
For example, in the code from Figure~\ref{fig:memcpy},
there are two instrumented bounds checks for each iteration of the loop
(one for the read and one for the write).
A more ``natural''/optimal approach is to check the bounds once for
each pointer outside of the loop, as shown in Figure~\ref{fig:memcpy2}.
Here we use \verb+lowfat_usable_size+ to determine the number of bytes
available in the \verb+src+ and \verb+dst+ buffers, and verify that this
is consistent with the parameter \verb+n+.
Such instrumentation can be added manually by the programmer, assuming
that objects are allocated using the \lowfat allocator.

In principle, the automatic instrumentation could be further optimized,
e.g., by using program analysis to automatically transform
Figure~\ref{fig:memcpy} into~\ref{fig:memcpy2}.
However, program analysis generally has limitations, and cannot
optimize all cases.
Furthermore, in some applications the programmer needs fine grained control
over what to instrument, in order to achieve an acceptable overhead versus
security ratio.
Thus, the programmer can restrict instrumentation to specific operations
(e.g., \verb+memcpy+) or specific pointers to sensitive data.

The overheads of manual bounds checking depend on how much is instrumented.

\subsubsection*{Bonus: Finding \texttt{free} API errors}

The \lowfat API can be also be
used to find some memory errors relating to \verb+free+.
For example, a stack or global
object should not be free'ed:
\begin{Verbatim}[samepage,commandchars=\\\{\},codes={\catcode`$=3\catcode`^=7}]
  if (lowfat_is_heap_ptr($ptr$)) lowfat_free($ptr$);
  else error();
\end{Verbatim}
In a similar vein, a pointer which is not the base of a heap object,
e.g. an interior heap pointer, should not be free'ed:
\begin{Verbatim}[samepage,commandchars=\\\{\},codes={\catcode`$=3\catcode`^=7}]
  if (lowfat_is_heap_ptr($ptr$) &&
      !lowfat_offset($ptr$)) lowfat_free($ptr$);
  else error();
\end{Verbatim}
We remark that general use-after-free checking is beyond the scope of the \lowfat API.
Testing if a pointer is free or not is known to suffer from races
(test versus usage) in multi-threaded environments.

\subsection{Conservative Garbage Collection}

Another application of the \lowfat allocator is for 
marking in conservative garabage collection for C/C++.
Under this idea, the \lowfat heap allocator itself is modified to
automatically invoke a mark-sweep collection phase eliminating the
need to manually free objects.
As is the standard approach, the ``mark'' phase scans for all objects
\emph{reachable} from some \emph{root set} of pointers, typically
global and stack memory.
Any reachable object is ``marked'' using internal meta-data associated
with each object.
Next, a ``sweep'' frees all unmarked (unreachable) objects since these
are no longer referenced by the program.
The garbage collector is \emph{conservative} meaning that it does not
rely on C/C++ type information --- rather any bit pattern that could be a
pointer is assumed to be a pointer.
The trade-offs for conservative collection are well known, e.g.,
see~\cite{boehm88garbage}.

The low-fat API can assist with marking algorithm as shown in
Figure~\ref{fig:mark}.
Here, given a potential pointer value \verb+ptr+, we first check if
\verb+ptr+ points to a heap object (lines 3--4).
Since \verb+ptr+ may be an \emph{interior pointer}, i.e., point
\emph{inside} an allocated object, we next retrieve the object's base
address by a call to \verb+lowfat_base+ (line 5).
We assume that \verb+set_mark+ marks the object in some disjoint meta-data
(lines 6--7),
and returns \verb+true+ if the object was already marked (to terminate loops).
Finally, we scan the object (lines 8--12)
and mark any bitpatten that happens to
be a valid pointer.
The disjoint meta-data itself is implemented as a collection of bitmaps
(one for each region, created using \verb+mmap+), with one bit for every
object within the corresponding region.

Note that the Boehm conservative garbage collector~\cite{boehm88garbage}
implements a
similar marking algorithm, but with its own implementations
of the size and base operations.
This is also one reason why the extended Boehm GC API is similar to
the \lowfat API.

\begin{figure}
\begin{lstlisting}[
    language=C,
    frame=single,
    numbers=left,
    numberstyle=\tiny,
    basicstyle=\scriptsize\ttfamily\fontseries{m}\selectfont,
    keywordstyle=\ttfamily\fontseries{b}\selectfont,
    mathescape]
void mark(void *ptr)
{
  void *base = lowfat_base(ptr);
  if (base == NULL)
    return; // Not low-fat
  if (set_mark(base))
    return; // Already marked.
  void **itr = (void **)base,
       **end = (void **)(base + lowfat_size(base));
  for (; itr < end; itr++)
    mark(*itr);
}
\end{lstlisting}
\caption{\lowfat API enhanced marking algorithm.\label{fig:mark}}
\end{figure}

\subsection{Hidden Meta-Data}

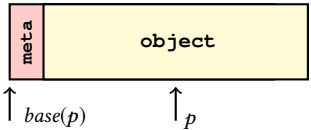
\begin{figure}
\begin{center}
{
\renewcommand{\ttdefault}{pcr}
\tikzstyle{memory}=[draw, rectangle, minimum height=1cm, minimum width=10em,
    anchor=west,thick]
\begin{tikzpicture}[y=-1cm, node distance=0 cm,outer sep = 0pt]
\node[memory,fill=red!20,text width=3.2cm] (meta) at (0,0) {\rotatebox{90}{\scriptsize \texttt{\textbf{meta}}} \hfill};
\node[memory,fill=yellow!20] (object) at (0.45,0) {\footnotesize ~~~~~~~~~~~~~~~~~~\texttt{\textbf{object}}~~~~~~~~~~~~~~~~~~~~~~~~};
\draw [->, memory] (0,1.05) -- ($ (meta.south west) + (0,0.1) $);
\node (base) at (0.6,1) {\footnotesize $\base{p}$};
\draw [->, memory] ($ (object) + (0,1.05) $) -- ($ (object.south) + (0,0.1) $);
\node (p) at ($ (object) + (0.2,1.05) $) {\footnotesize $p$};
\end{tikzpicture}
}
\end{center}
\caption{(Hidden) meta-data stored at the base of an
    object.\label{fig:metadata}}
\end{figure}

The \lowfat API can also be used to associate arbitrary meta-data to allocated
objects.
The basic idea is to store the meta-data at the base of the object, as
illustrated in Figure~\ref{fig:metadata}.
Here $p$ is a (possibly interior) pointer to a \lowfat allocated
(\texttt{object}), and the meta-data (\texttt{meta}) is stored
at the base of the allocation.
The meta-data can be transparently bound to an object by wrapping memory
allocation, such as the following:
\begin{Verbatim}[fontsize=\small,samepage,commandchars=\\\{\},codes={\catcode`$=3\catcode`^=7}]
  void *meta_malloc(size_t $\mathit{size}$, META $m$) \{
    META *ptr = lowfat_malloc($\mathit{size}$ + sizeof(META));
    *ptr = $m$;
    return (ptr + 1);
  \}
\end{Verbatim}
Note the function returns (\verb_ptr + 1_), meaning that the meta-data
is hidden from the program, analogous to a hidden \verb+malloc+ header
that occupies the memory immediately before the allocated object.
However, a crucial difference with \verb+malloc+ headers is that
in \verb+malloc+ accessing the header is restricted through a base pointer, 
here, we have no restrictions. 
Later, the meta-data can be retrieved via a call to \verb+lowfat_base+,
as follows:
\begin{Verbatim}[commandchars=\\\{\},codes={\catcode`$=3\catcode`^=7}]
    $\mathit{m}$ = *(META *)lowfat_base($\mathit{p}$);
\end{Verbatim}
The same basic idea can be extended to both stack and global objects,
but requires a compiler transformation.
Stack allocation is transformed in a similar way to \verb+malloc+, where
\begin{Verbatim}[fontsize=\small,samepage,commandchars=\\\{\},codes={\catcode`$=3\catcode`^=7}]
    $\mathit{ptr}$ = alloca($\mathit{size}$);
\end{Verbatim}
is transformed into:
\begin{Verbatim}[fontsize=\small,samepage,commandchars=\\\{\},codes={\catcode`$=3\catcode`^=7}]
    META *mptr = lowfat_alloca($\mathit{size}$ + sizeof(META));
    *mptr = $m$;
    $\mathit{ptr}$ = (mptr + 1);
\end{Verbatim}
Here, \verb+lowfat_alloca+ is itself expanded via program transformation,
as per~\cite{duck17stack}.
We note that the usage of \verb+alloca+ is just for the sake of an example,
and the transform is applicable to all forms of stack allocation.
In particular, the use of \verb+alloca+ can be internal to the compiler
as is the case with LLVM.

Globals are more difficult to transform, since a global is also a
symbol that may be referenced externally, possibly by code not subject to
the automatic program transformation.
Thus, we cannot rely on solutions that change the \emph{Application Binary
Interface} (ABI).
To fix this, we use a simple \emph{symbol-within-a-symbol} trick.
The basic idea is as follows: given the original global variable definition:
\begin{Verbatim}[samepage,commandchars=\\\{\},codes={\catcode`$=3\catcode`^=7}]
  T $\mathit{global}$ = $\mathit{definition}$;
\end{Verbatim}
We first define a \emph{wrapper type} of the form:
\begin{Verbatim}[samepage,commandchars=\\\{\},codes={\catcode`$=3\catcode`^=7}]
  struct wrapper \{ META m; T data; \};
\end{Verbatim}
We also ensure that the structure is \emph{packed} (e.g. by using
the GCC \verb+packed+ attribute), meaning that there will be no gap between the
\verb+m+ and \verb+data+ fields.
Next, we replace the original global with the wrapped version
\begin{Verbatim}[samepage,commandchars=\\\{\},codes={\catcode`$=3\catcode`^=7}]
  struct wrapper $\mathit{wrappedGlobal}$ = \{$m$, $\mathit{definition}$\};
\end{Verbatim}
The program (including external modules) may still reference the
original $\mathit{global}$ symbol.
To fix this we define $\mathit{global}$ to point to the \verb+data+
field inside $\mathit{wrappedGlobal}$.
The most direct way to do this is via (module-level) inline assembly:
\begin{Verbatim}[samepage,commandchars=\\\{\},codes={\catcode`$=3\catcode`^=7}]
  asm (".globl $\mathit{global}$"
       ".set $\mathit{global}$, $\mathit{wrappedGlobal}$+$\mathit{size}$");
\end{Verbatim}
where $\mathit{size}{=}\mathtt{sizeof}(\mathtt{META})$.
By using this symbol-within-a-symbol trick, the global variable
($\mathit{global}$) can be used as normal by the program, including
by external untransformed modules.

A form of the hidden meta-data approach is 
used by EffectiveSan~\cite{duck18effective} to
store object dynamic type information, a.k.a., the \emph{effective} type
of allocated objects, in order to support dynamic type checking for
C/C++.
However, there is no limit on the kinds of meta-data that can be stored.
Like other generic meta-data storage schemes, such as
\emph{Padding Area MetaData} (PAMD), there exist many
other potential applications, including accurate (exact object size)
bounds-checking, profiling and statistics, flow tracking, and data race
detection~\cite{liu17metadata}.
METAlloc~\cite{haller16meta} is another general meta data framework,
but uses its own
shadow memory scheme, and is quite different to our approach and PAMD.

\subsection{Typed Pointers}

\begin{figure}
\centering
\subfloat[Header Pointers\label{fig:header}]{
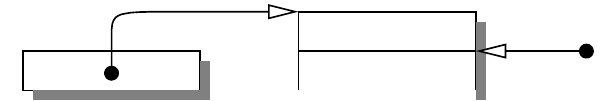
} \\
\subfloat[Tagged Pointers\label{fig:tagged}]{
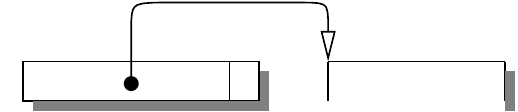
} \\
\subfloat[Partitioned Pointers\label{fig:part}]{
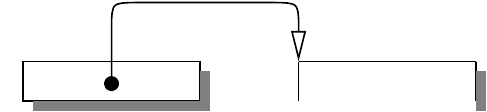
}
\caption{Common typed-pointer representations.\label{fig:typed_pointers}}
\end{figure}

A \emph{typed pointer} 
is one of various methods for associating
dynamic type information with pointers.
There are several existing methods~\cite{gudeman93type} for associating a
type $t$ to a pointer $p$ to form a typed-pointer $q$.
These include:
\begin{itemize}[leftmargin=*]
\item[-] \emph{Headers}: store $t$ within the object pointed to by $p$
(Figure~\ref{fig:header});
\item[-] \emph{Tagged}: fold $t$ into the representation of $p$ itself
(Figure~\ref{fig:tagged});
\item[-] \emph{Partitioned}: allocate $p$ from different regions based on $t$
(Figure~\ref{fig:part}).
\end{itemize}
Each approach as its own advantages/disadvantages:
\emph{header pointers} is portable but consumes memory to store $t$;
\emph{tagged pointers} and \emph{partitioned pointers} do not consume
more memory, but rely on knowledge about the underlying memory management
system.

In this section we explore some alternatives/extensions based on the
\lowfat API, namely:
\emph{size-typed} pointers and \emph{extended tagged} pointers.

\subsubsection*{Size-typed pointers}

One idea is to distinguish pointer types based on the allocation size, a.k.a.
\emph{size-typed pointers}.
The size can be determined very quickly via the \verb+lowfat_index+ API call,
however, this approach is only applicable to objects where each supported
dynamic type happens to correspond to a different allocation size.
That said, real-world applications exist, as illustrated by the
following example:
\begin{example}[2-3-4 Trees]\label{ex:234}
To illustrate size-typed pointers we consider an implementation 
of \emph{2-3-4 tree}s~\cite{algorithms11sedgewick}.
A 2-3-4 tree is a self-balancing tree data-structure that can be
used to implement \emph{associative arrays} mapping keys to values.
For example, the following
\begin{center}
\begin{picture}(0,0)%
\includegraphics{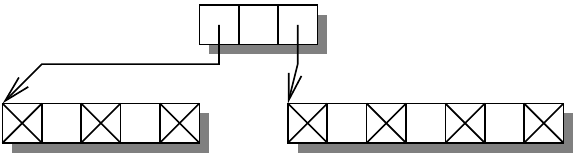}%
\end{picture}%
\setlength{\unitlength}{4144sp}%
\begingroup\makeatletter\ifx\SetFigFont\undefined%
\gdef\SetFigFont#1#2#3#4#5{%
  \reset@font\fontsize{#1}{#2pt}%
  \fontfamily{#3}\fontseries{#4}\fontshape{#5}%
  \selectfont}%
\fi\endgroup%
\begin{picture}(2623,688)(529,-827)
\put(1688,-286){\makebox(0,0)[lb]{\smash{{\SetFigFont{8}{9.6}{\rmdefault}{\mddefault}{\updefault}{\color[rgb]{0,0,0}$5$}%
}}}}
\put(788,-738){\makebox(0,0)[lb]{\smash{{\SetFigFont{8}{9.6}{\rmdefault}{\mddefault}{\updefault}{\color[rgb]{0,0,0}$1$}%
}}}}
\put(1149,-736){\makebox(0,0)[lb]{\smash{{\SetFigFont{8}{9.6}{\rmdefault}{\mddefault}{\updefault}{\color[rgb]{0,0,0}$2$}%
}}}}
\put(2454,-739){\makebox(0,0)[lb]{\smash{{\SetFigFont{8}{9.6}{\rmdefault}{\mddefault}{\updefault}{\color[rgb]{0,0,0}$8$}%
}}}}
\put(2816,-736){\makebox(0,0)[lb]{\smash{{\SetFigFont{8}{9.6}{\rmdefault}{\mddefault}{\updefault}{\color[rgb]{0,0,0}$9$}%
}}}}
\put(2093,-738){\makebox(0,0)[lb]{\smash{{\SetFigFont{8}{9.6}{\rmdefault}{\mddefault}{\updefault}{\color[rgb]{0,0,0}$6$}%
}}}}
\end{picture}%

\end{center}
is a 2-3-4 tree consisting of a root 2-node, a left child 3-node, and a
right child 4-node.
The name ``2-3-4'' represents the three node types:
\emph{2-node}s, \emph{3-node}s, and \emph{4-node}s, which are of sizes
(in 8byte words) of 3, 5, and 7 respectively.
This means the nodes will be allocated from different region \#2, \#3, and \#5
respectively, assuming the standard size configuration.
Thus, given a pointer \verb+ptr+ to a (undetermined) 2-3-4 node, we can
efficiently determine the dynamic type by using the \verb+lowfat_index+
operation.
$\qed$\end{example}
Size-typed pointers are essentially a special case of partitioned pointers.
The main advantage is that the \lowfat allocator supports the functionality
directly, rather than requiring the programmer to implement a
specialized allocator.

\subsubsection*{Extended Tagged Pointers}

Sized-typed pointers have limited applicability, since the mapping from
types to allocation sizes must be one-to-one.
Tagged pointers are more general, but the number of tag bits can be limited.
For this, we introduce the notion of \emph{extended tagged pointers} which
are a generalization of standard tagged pointers using the unused lower
$N$-bits (typically $N{=}4$) of allocated objects.
Assuming $N{=}4$ this allows for 16 distinct types, whereas extended tagged
pointers can store up to $\mathit{size}$ distinct types, where
$\mathit{size}$ is the allocation size of the object.
Normally, for standard tagged pointers, the type ($\mathit{tag}$)
can be retrieved via a simple bitmask operation, e.g.,
\begin{Verbatim}[commandchars=\\\{\},codes={\catcode`$=3\catcode`^=7}]
    $\mathit{tag}$ = $\textttz{ptr}$ & 0xF
\end{Verbatim}
However, using the \lowfat API, we can generalize this as follows:
\begin{Verbatim}[commandchars=\\\{\},codes={\catcode`$=3\catcode`^=7}]
    $\mathit{tag}$ = lowfat_offset($\mathit{ptr}$)
\end{Verbatim}
This supports all possible tag values within the range
$[0..\mathit{size})$.
Alternatively, tagged pointers may use the unused high bits
(typically 16 bits for \verb+x86_64+).
Extended tagged pointers may replace or be used in conjunction with high tag
bits, depending the application.

The \verb+lowfat_offset+
operation is generally slower than the constant bitmask operation
required standard tagged pointers, especially if fixed-point arithmetic is
used.
Thus, there exists trade-off between performance and number of types,
meaning the usefulness is application dependent.
We provide one such application in Section~\ref{sec:vectors}.

\subsubsection*{Evaluation: 2-3-4 trees}

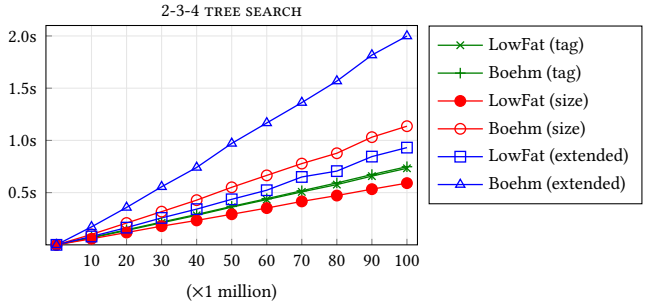
\begin{figure}
\centering
\pgfplotsset{
    axisA/.style={
        xmin=0,
        xmax=10,
        ymin=0,
        ymax=2100,
        ytick={500,1000,1500,2000},
        yticklabels={
            {\scriptsize 0.5s},
            {\scriptsize 1.0s},
            {\scriptsize 1.5s},
            {\scriptsize 2.0s},
        },
        xticklabels={
            {\scriptsize 10},
            {\scriptsize 20},
            {\scriptsize 30},
            {\scriptsize 40},
            {\scriptsize 50},
            {\scriptsize 60},
            {\scriptsize 70},
            {\scriptsize 80},
            {\scriptsize 90},
            {\scriptsize 100}
        },
        xlabel near ticks,
        xlabel={\scriptsize ($\times1$ million)},
        scaled y ticks = false,
        width=0.77\columnwidth,
        height=4.5cm,
        xtick={1,2,3,4,5,6,7,8,9,10},
        xtick pos=left,
        ytick pos=left,
        legend cell align={left},
        major tick length=0.08cm,
        enlarge x limits={true, abs value=0.3},
        grid style={gray!20},
        grid=both,
        legend pos=outer north east,
        title style={yshift=-6pt}
        }
}
\pgfplotstableread{
0  0   0   0   0    0   0 
1  67  72  58  100  80  172     
2  138 144 118 209  165 358 
3  211 218 179 318  258 556 
4  283 291 234 428  342 741 
5  361 368 293 551  436 971 
6  433 442 352 665  523 1167 
7  507 518 416 778  650 1362 
8  577 593 472 877  706 1568 
9  658 673 533 1031 845 1817 
10 733 748 591 1136 931 1999 
}\dataset
\begin{tikzpicture}
    \begin{axis}[
        axisA,
        title={\scriptsize\sc 2-3-4 tree search}
      ]
      \addplot[mark options={solid},line width=0.5,mark=x,mark size=2,color=mygreen] table[x index=0,y index=1] \dataset;
      \addplot[mark options={solid},line width=0.5,mark=+,mark size=2,color=mygreen] table[x index=0,y index=2] \dataset;
      \addplot[mark options={solid},line width=0.5,mark=*,mark size=2,color=red] table[x index=0,y index=3] \dataset;
      \addplot[mark options={solid},line width=0.5,mark=o,mark size=2,color=red] table[x index=0,y index=4] \dataset;
      \addplot[mark options={solid},line width=0.5,mark=square,mark size=2,color=blue] table[x index=0,y index=5] \dataset;
      \addplot[mark options={solid},line width=0.5,mark=triangle,mark size=2,color=blue] table[x index=0,y index=6] \dataset;
      \legend{{\scriptsize \lowfat (tag)},{\scriptsize Boehm (tag)},
        {\scriptsize \lowfat (size)},{\scriptsize Boehm (size)},
        {\scriptsize \lowfat (extended)},{\scriptsize Boehm (extended)}}
    \end{axis}
\end{tikzpicture}
\caption{2-3-4 tree typed pointer performance results.\label{fig:234}}
\end{figure}

We evaluate both size-typed and extended tagged pointers for 2-3-4 trees.
Our benchmark consists of a searching for every key in a 2-3-4 tree of
size $N$, measured in seconds.
We compare six different versions:
a standard tagged pointer implementation (\verb+tag+) using the lower
4 tag bits,
an implementation using size-typed pointers (\verb+size+), and
an implementation using extended tagged pointers (\verb+extended+).
Although extended tagged pointers are overkill for 2-3-4 trees,
it is nevertheless a useful test for performance evaluation.
We compare each version implemented either the \lowfat API, or
using the similar Boehm GC API.
For the Boehm tests, we use manual memory management mode.

The results are shown in Figure~\ref{fig:234}.
Unsurprisingly, the (\verb+tag+) tests (which do not use any special API calls)
show little difference in performance between the two versions.
For \lowfat, size-typed pointers (\verb+size+) are even faster than
traditional tagged pointers by $\sim$20\%.
This shows that size-typed pointers are a good alternative for performance
critical code, under the caveat that size-typing is applicable to target
data-structure.
Extended tagged pointers (\verb+extended+) are slower than traditional
tagged pointers by $\sim$27\%, so should only be used for applications
that require extra tag bits.
Also unsurprisingly, the Boehm variants of size-typed and extended tagged
pointers were much slower than the \lowfat version, e.g.
${>}{\times}2$ for extended tagged pointers.
This is because the \lowfat API is highly optimized and inlined
for the size/base operations, whereas the Boehm API requires library calls.

\subsection{Low-fat Vectors}\label{sec:vectors}

\begin{figure}
\centering
\pgfplotsset{
    axisA/.style={
        xmin=1,
        xmax=100,
        ymin=0,
        ymax=7000000,
        ytick={1000000,2000000,3000000,4000000,5000000,6000000,7000000},
        yticklabels={
            {\scriptsize 1s},
            {\scriptsize 2s},
            {\scriptsize 3s},
            {\scriptsize 4s},
            {\scriptsize 5s},
            {\scriptsize 6s},
            {\scriptsize 7s}
        },
        xticklabels={
            {\scriptsize 10},
            {\scriptsize 20},
            {\scriptsize 30},
            {\scriptsize 40},
            {\scriptsize 50},
            {\scriptsize 60},
            {\scriptsize 70},
            {\scriptsize 80},
            {\scriptsize 90},
            {\scriptsize 100}
        },
        xlabel near ticks,
        xlabel={\scriptsize ($\times10$ million)},
        scaled y ticks = false,
        width=0.77\columnwidth,
        height=4.5cm,
        xtick={10,20,30,40,50,60,70,80,90,100},
        xtick pos=left,
        ytick pos=left,
        legend cell align={left},
        major tick length=0.08cm,
        enlarge x limits={true, abs value=0.3},
        grid style={gray!20},
        grid=both,
        legend pos=outer north east,
        title style={yshift=-6pt}
        }
}
\pgfplotstableread{
0    0        0       0
5    311841   213540  147496
10   653553   456458  288215
15   939181   644727  433566
20  1224347   830541  571393
25  1511535  1016194  708740
30  2019668  1424915  859204
35  2305763  1610207  996144
40  2593476  1795223 1133844
45  2879462  1980033 1271375
50  3165547  2165506 1409707
55  3451504  2351315 1575523
60  3737374  2537066 1714020
65  4022220  2722204 1853364
70  4307977  2907335 1992299
75  4594457  3093579 2131056
80  4879937  3275480 2270494
85  5165602  3466511 2409469
90  5451265  3654626 2548545
95  5737478  3843199 2687640
100 6022468  4031239 2827002
}\dataset
\begin{tikzpicture}
    \begin{axis}[
        axisA,
        title={\scriptsize\sc vector construction}
      ]
      \addplot[mark options={solid},line width=0.5,mark=x,mark size=2,color=mygreen] table[x index=0,y index=1] \dataset;
      \addplot[mark options={solid},line width=0.5,mark=+,mark size=2,color=mygreen] table[x index=0,y index=2] \dataset;
      \addplot[line width=0.5,mark=triangle,mark size=2,color=red!80] table[x index=0,y index=3] \dataset;
      \legend{{\scriptsize \lowfat},{\scriptsize \lowfat Pow2},
          {\scriptsize Fat}}
    \end{axis}
\end{tikzpicture}
\pgfplotstableread{
0    0        0       0
5   97387    98095    81094     
10  194619   196104   162566
15  292016   294405   243874
20  389119   391895   324701
25  486222   490382   405594
30  584893   588588   486507
35  682938   686102   567821
40  781245   784137   648766
45  883865   881627   729672
50  981550   978646   810665
55  1078958  1076826  891900
60  1177172  1174424  972762
65  1274715  1271926  1053631
70  1372046  1368959  1134802
75  1469156  1466324  1216065
80  1566451  1563679  1296903
85  1664447  1661808  1377744
90  1761722  1760203  1458588
95  1860923  1858139  1539746
100 1958165  1956915  1620707
}\dataset
\begin{tikzpicture}
    \begin{axis}[
        axisA,
        title={\scriptsize\sc vector access}
      ]
      \addplot[mark options={solid},line width=0.5,mark=x,mark size=2,color=mygreen] table[x index=0,y index=1] \dataset;
      \addplot[mark options={solid},line width=0.5,mark=+,mark size=2,color=mygreen] table[x index=0,y index=2] \dataset;
      \addplot[line width=0.5,mark=triangle,mark size=2,color=red!80] table[x index=0,y index=3] \dataset;
      \legend{{\scriptsize \lowfat},{\scriptsize \lowfat Pow2},
          {\scriptsize Fat}}
    \end{axis}
\end{tikzpicture}
\caption{Low-fat vector benchmarks in seconds.\label{fig:vecbench}}
\vspace{-2mm}
\end{figure}
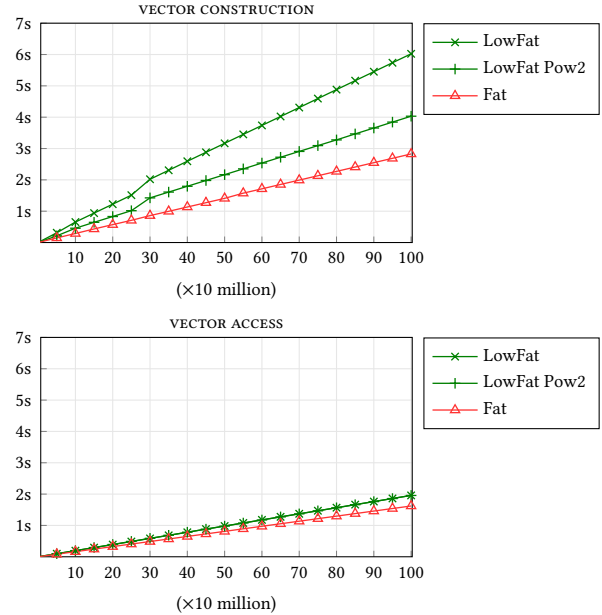

A very common data-structure is a \emph{vector}, for example \verb_C++_'s
\verb+std::vector+, which typically
consists of three core components:
an array of items \texttt{data} (vector data), a length \texttt{len} (vector
length), and a current position \texttt{pos} (next free item).
Vectors are normally implemented as structures containing these
three components:
\begin{verbatim}
struct vector {size_t len;
               size_t pos;
               item *data;}
\end{verbatim}
We refer to such representations as ``fat'' vectors.

Using the \lowfat API we can implement a more compact representation, a.k.a.
``low-fat'' vectors.
For this, we define a vector to be an array of items:
($\texttt{typedef}~\mathit{item}~\texttt{*}\mathit{vector}$).
The $\mathit{len}$ field becomes implicit, and can be calculated
dynamically using \verb+lowfat_size+:
\begin{Verbatim}[commandchars=\\\{\},codes={\catcode`$=3\catcode`^=7}]
    $\mathit{len}$ = lowfat_size($\mathit{vector}$) / sizeof($\mathit{item}$)
\end{Verbatim}
The $\mathit{pos}$ can be stored as an \emph{extended tag}, i.e.
\begin{Verbatim}[commandchars=\\\{\},codes={\catcode`$=3\catcode`^=7}]
    $\mathit{pos}$ = lowfat_offset($\mathit{vector}$)
    $\mathit{data}$ = lowfat_base($\mathit{vector}$)
\end{Verbatim}

\subsubsection*{Evaluation: Low-fat vectors}

The main advantage of low-fat vectors is that they eliminate the 
need to explicitly store the \verb+len+, \verb+pos+ and \verb+data+ fields.
Assuming that \verb+len+, \verb+pos+, (\verb+item *+) and \verb+item+
are all 1-word in size, then if a fat vector consumes
$n$ words, the corresponding low-fat vector will consume $(n-3)$ words.
The trade-off is that (re)calculating fields incurs
additional overheads compared to storing the values directly.
To evaluate the performance of low-fat vectors, we benchmark constructing
a single vector of integers using the \verb+push_back+ operation.
Next, we evaluate the time taken to calculate the sum of all elements of
the vector.
The results are shown in Figure~\ref{fig:vecbench} illustrating the
classic space-time tradeoff.
We see that constructing low-fat vectors is ${\sim}2\times$ overhead
for non-power-of-two sizes, ${\sim}1.33\times$ overhead for power-of-two sizes.
Reading from low-fat vector incurs a ${\sim}1.2\times$ overhead for
both versions.
Thus, low-fat vectors are best suited for programs that create large 
numbers of small vectors and where optimizing memory overheads are the priority.

\section{Conclusions}\label{sec:conclusion}

In this paper we presented an extended \lowfat memory allocation API.
The main advantage of the \lowfat API extensions is that some operations,
namely, finding the size/base/offset of pointers, relative to the
original allocation, are very fast operations
(typically can be implemented in a few inlined instructions).
We argue that these properties enable several applications for the
\lowfat allocator that are not feasible with existing allocators,
such as bounds checking, generic meta-data storage, typed pointers
and compact data-structures.
We evaluated several of these ideas, with promising results.
The \verb+malloc+ API has been essentially unchanged for a long time,
we believe that the idea of memory allocation API extensions going beyond
the core allocator function is a genuinely useful and practical addition.

\section*{Acknowledgements}

This research was partially supported by a grant from the
National Research Foundation,
Prime Minister's Office, Singapore under its National
Cybersecurity R\&D Program (TSU\-NA\-Mi project, No.
NRF2014NCR-NCR001-21)
and administered by the National Cybersecurity R\&D Directorate.

\bibliographystyle{ACM-Reference-Format}
\bibliography{lowfat}


\begin{thebibliography}{18}


\ifx \showCODEN    \undefined \def \showCODEN     #1{\unskip}     \fi
\ifx \showDOI      \undefined \def \showDOI       #1{#1}\fi
\ifx \showISBNx    \undefined \def \showISBNx     #1{\unskip}     \fi
\ifx \showISBNxiii \undefined \def \showISBNxiii  #1{\unskip}     \fi
\ifx \showISSN     \undefined \def \showISSN      #1{\unskip}     \fi
\ifx \showLCCN     \undefined \def \showLCCN      #1{\unskip}     \fi
\ifx \shownote     \undefined \def \shownote      #1{#1}          \fi
\ifx \showarticletitle \undefined \def \showarticletitle #1{#1}   \fi
\ifx \showURL      \undefined \def \showURL       {\relax}        \fi
\providecommand\bibfield[2]{#2}
\providecommand\bibinfo[2]{#2}
\providecommand\natexlab[1]{#1}
\providecommand\showeprint[2][]{arXiv:#2}

\bibitem[\protect\citeauthoryear{??}{je-}{2018}]%
        {je-malloc}
 \bibinfo{year}{2018}\natexlab{}.
\newblock \bibinfo{title}{jemalloc memory allocator}.
\newblock
\newblock
\showURL{%
\url{http://jemalloc.net/}}


\bibitem[\protect\citeauthoryear{??}{lea}{2018}]%
        {lea-malloc}
 \bibinfo{year}{2018}\natexlab{}.
\newblock \bibinfo{title}{Lea Memory Allocator}.
\newblock
\newblock
\showURL{%
\url{http://g.oswego.edu/dl/html/malloc.html}}


\bibitem[\protect\citeauthoryear{??}{low}{2018}]%
        {lowfat_github}
 \bibinfo{year}{2018}\natexlab{}.
\newblock \bibinfo{title}{LowFat: Lean C/C++ Bounds Checking with Low-Fat
  Pointers}.
\newblock
\newblock
\showURL{%
\url{https://github.com/GJDuck/LowFat}}


\bibitem[\protect\citeauthoryear{??}{tcm}{2018}]%
        {tcmalloc}
 \bibinfo{year}{2018}\natexlab{}.
\newblock \bibinfo{title}{TCMalloc: Thread-Caching Malloc}.
\newblock
\newblock
\showURL{%
\url{http://goog-perftools.sourceforge.net/doc/tcmalloc.html}}


\bibitem[\protect\citeauthoryear{Berger, Zorn, and McKinley}{Berger
  et~al\mbox{.}}{2001}]%
        {Berger2001}
\bibfield{author}{\bibinfo{person}{E. Berger}, \bibinfo{person}{B. Zorn}, {and}
  \bibinfo{person}{K. McKinley}.} \bibinfo{year}{2001}\natexlab{}.
\newblock \showarticletitle{Composing {H}igh-performance {M}emory
  {A}llocators}. In \bibinfo{booktitle}{{\em Programming Language Design and
  Implementation}}. \bibinfo{publisher}{ACM}.
\newblock


\bibitem[\protect\citeauthoryear{Boehm and Weiser}{Boehm and Weiser}{1988}]%
        {boehm88garbage}
\bibfield{author}{\bibinfo{person}{H. Boehm} {and} \bibinfo{person}{M.
  Weiser}.} \bibinfo{year}{1988}\natexlab{}.
\newblock \showarticletitle{Garbage collection in an uncooperative
  environment}.
\newblock \bibinfo{journal}{{\it Software Prac. Experience}}
  \bibinfo{volume}{18}, \bibinfo{number}{9} (\bibinfo{date}{Sept.}
  \bibinfo{year}{1988}), \bibinfo{pages}{807--820}.
\newblock


\bibitem[\protect\citeauthoryear{Dang, Maniatis, and Wagner}{Dang
  et~al\mbox{.}}{2015}]%
        {dang15shadow}
\bibfield{author}{\bibinfo{person}{T. Dang}, \bibinfo{person}{P. Maniatis},
  {and} \bibinfo{person}{D. Wagner}.} \bibinfo{year}{2015}\natexlab{}.
\newblock \showarticletitle{The {P}erformance {C}ost of {S}hadow {S}tacks and
  {S}tack {C}anaries}. In \bibinfo{booktitle}{{\em ACM Symposium on
  Information, Computer and Communications Security}}.
  \bibinfo{publisher}{ACM}.
\newblock


\bibitem[\protect\citeauthoryear{Duck and Yap}{Duck and Yap}{2016}]%
        {duck16heap}
\bibfield{author}{\bibinfo{person}{G. Duck} {and} \bibinfo{person}{R. Yap}.}
  \bibinfo{year}{2016}\natexlab{}.
\newblock \showarticletitle{{H}eap {B}ounds {P}rotection with {L}ow {F}at
  {P}ointers}. In \bibinfo{booktitle}{{\em Compiler Construction}}.
  \bibinfo{publisher}{ACM}.
\newblock


\bibitem[\protect\citeauthoryear{Duck and Yap}{Duck and Yap}{2018}]%
        {duck18effective}
\bibfield{author}{\bibinfo{person}{G. Duck} {and} \bibinfo{person}{R. Yap}.}
  \bibinfo{year}{2018}\natexlab{}.
\newblock \showarticletitle{Effective{S}an: {T}ype and {M}emory {E}rror
  {D}etection using {D}ynamically {T}yped {C}/{C}++}. In
  \bibinfo{booktitle}{{\em Programming Language Design and Implementation}}.
  \bibinfo{publisher}{ACM}.
\newblock


\bibitem[\protect\citeauthoryear{Duck, Yap, and Cavallaro}{Duck
  et~al\mbox{.}}{2017}]%
        {duck17stack}
\bibfield{author}{\bibinfo{person}{G. Duck}, \bibinfo{person}{R. Yap}, {and}
  \bibinfo{person}{L. Cavallaro}.} \bibinfo{year}{2017}\natexlab{}.
\newblock \showarticletitle{{S}tack {B}ounds {P}rotection with {L}ow {F}at
  {P}ointers}. In \bibinfo{booktitle}{{\em Network and Distributed System
  Security Symposium}}. \bibinfo{publisher}{The Internet Society}.
\newblock


\bibitem[\protect\citeauthoryear{Gudeman}{Gudeman}{1993}]%
        {gudeman93type}
\bibfield{author}{\bibinfo{person}{D. Gudeman}.}
  \bibinfo{year}{1993}\natexlab{}.
\newblock \bibinfo{title}{Representing {T}ype {I}nformation in {D}ynamically
  {T}yped {L}anguages}.
\newblock
\newblock
\newblock
\shownote{Technical Report.}


\bibitem[\protect\citeauthoryear{Haller, Kouwe, Giuffrida, and Bos}{Haller
  et~al\mbox{.}}{2016}]%
        {haller16meta}
\bibfield{author}{\bibinfo{person}{I. Haller}, \bibinfo{person}{E. Kouwe},
  \bibinfo{person}{C. Giuffrida}, {and} \bibinfo{person}{H. Bos}.}
  \bibinfo{year}{2016}\natexlab{}.
\newblock \showarticletitle{{META}lloc: {E}fficient and {C}omprehensive
  {M}etadata {M}anagement for {S}oftware {S}ecurity {H}ardening}. In
  \bibinfo{booktitle}{{\em European Workshop on System Security}}.
  \bibinfo{publisher}{ACM}.
\newblock


\bibitem[\protect\citeauthoryear{{Intel Corporation}}{{Intel
  Corporation}}{2018}]%
        {intel}
\bibfield{author}{\bibinfo{person}{{Intel Corporation}}.}
  \bibinfo{year}{2018}\natexlab{}.
\newblock \bibinfo{title}{{Intel 64 and IA-32 Architectures Optimization
  Reference Manual}}.
\newblock
\newblock


\bibitem[\protect\citeauthoryear{Kwon, Dhawan, Smith, Knight, and DeHon}{Kwon
  et~al\mbox{.}}{2013}]%
        {kwon13lowfat}
\bibfield{author}{\bibinfo{person}{A. Kwon}, \bibinfo{person}{U. Dhawan},
  \bibinfo{person}{J. Smith}, \bibinfo{person}{T. Knight}, {and}
  \bibinfo{person}{A. DeHon}.} \bibinfo{year}{2013}\natexlab{}.
\newblock \showarticletitle{Low-fat {P}ointers: {C}ompact {E}ncoding and
  {E}fficient {G}ate-level {I}mplementation of {F}at {P}ointers for {S}patial
  {S}afety and {C}apability-based {S}ecurity}. In \bibinfo{booktitle}{{\em
  Computer and Communications Security}}. \bibinfo{publisher}{ACM}.
\newblock


\bibitem[\protect\citeauthoryear{Liu and Criswell}{Liu and Criswell}{2017}]%
        {liu17metadata}
\bibfield{author}{\bibinfo{person}{Z. Liu} {and} \bibinfo{person}{J.
  Criswell}.} \bibinfo{year}{2017}\natexlab{}.
\newblock \showarticletitle{Flexible and {E}fficient {M}emory {O}bject
  {M}etadata}. In \bibinfo{booktitle}{{\em International Symposium on Memory
  Management}}. \bibinfo{publisher}{ACM}.
\newblock


\bibitem[\protect\citeauthoryear{LLVM}{LLVM}{2018}]%
        {llvm}
LLVM \bibinfo{year}{2018}\natexlab{}.
\newblock \bibinfo{howpublished}{\texttt{http://llvm.org}}.
\newblock


\bibitem[\protect\citeauthoryear{Sedgewick and Wayne}{Sedgewick and
  Wayne}{2011}]%
        {algorithms11sedgewick}
\bibfield{author}{\bibinfo{person}{R. Sedgewick} {and} \bibinfo{person}{K.
  Wayne}.} \bibinfo{year}{2011}\natexlab{}.
\newblock \bibinfo{booktitle}{{\em Algorithms\/} (\bibinfo{edition}{4th} ed.)}.
\newblock \bibinfo{publisher}{Addison-Wesley Professional}.
\newblock


\bibitem[\protect\citeauthoryear{Wilson, Johnstone, Neely, and Boles}{Wilson
  et~al\mbox{.}}{1995}]%
        {wilson-survey}
\bibfield{author}{\bibinfo{person}{P. Wilson}, \bibinfo{person}{M. Johnstone},
  \bibinfo{person}{M. Neely}, {and} \bibinfo{person}{D. Boles}.}
  \bibinfo{year}{1995}\natexlab{}.
\newblock \bibinfo{booktitle}{{\em Dynamic {S}torage {A}llocation: a {S}urvey
  and {C}ritical {R}eview}}.
\newblock \bibinfo{publisher}{Springer}.
\newblock


\end{thebibliography}

\appendix

\section{Low-fat Parameters}\label{app:params}

\newcommand{\set}[1]{\mathsf{#1}}

{\small
\begin{align*}
& \mathtt{LOWFAT\_REGION\_SIZE} = 32\mathit{GB} \\
& M = |\set{Sizes}| = 61 \\
& \set{Sizes} =
\begin{array}{l}
\langle
16,
32,
48,
64,
80,
96,
112,
128,
144,
160,
192,
224,
256, \\
~272,
320,
384,
448,
512,
528,
640,
768,
896,
1024, 
1040, \\
~1280,
1536,
1792,
2048, 
2064,
2560,
3072,
3584,
4096, \\
~4112,
5120,
6144,
7168,
8192, 
8208,
10240,
12288, \\
~16\mathit{KB},
32\mathit{KB},
64\mathit{KB},
128\mathit{KB},
256\mathit{KB},
512\mathit{KB}, 
1\mathit{MB}, \\
~2\mathit{MB},
4\mathit{MB},
8\mathit{MB},
16\mathit{MB}, 
32\mathit{MB},
64\mathit{MB},
128\mathit{MB}, \\
~256\mathit{MB},
512\mathit{MB},
1\mathit{GB},
2\mathit{GB},
4\mathit{GB},
8\mathit{GB}
\rangle
\end{array} 
\end{align*}
}

\end{document}